\documentstyle[aps,prl,twocolumn,epsfig]{revtex}

\begin{document}

\title{Interferometric detection of a single vortex in a dilute 
Bose-Einstein condensate}

\author{F. Chevy, K. W. Madison, V. Bretin, and J. Dalibard}
\address{Laboratoire Kastler Brossel\cite{byline}, D\'epartement de 
Physique de
l'Ecole Normale Sup\'erieure,
24 rue Lhomond, 75005 Paris, France}
\date{\today}

\twocolumn[\hsize\textwidth\columnwidth\hsize\csname@twocolumnfalse\endcsname

\maketitle

\begin{abstract}
Using two radio frequency pulses separated in time
we perform an amplitude division interference experiment
on a rubidium Bose-Einstein condensate.  The presence of a
quantized vortex, which is nucleated by stirring the 
condensate with a laser beam, is revealed by a dislocation
in the fringe pattern.
\end{abstract}

\pacs{PACS numbers: 03.75.Fi, 67.40.Db, 32.80.Lg}


\vskip1pc
] 

The experimental realization of Bose Einstein condensation of 
dilute atomic gases has opened for study a whole new class 
of macroscopic quantum systems \cite{Varenna98}. 
In particular, the question of the superfluidity of a dilute 
condensate can now be addressed experimentally 
by analyzing the response of such a system to a moving perturbation.
This response is severely restricted by the strong constraints that 
exist on the velocity field of the condensate. For a macroscopic wave function 
$\psi({\bf r})=\sqrt{n({\bf r})}\,e^{i S({\bf r})}$,
where $n({\bf r})$ is the density of the condensate and $S({\bf r})$
the local phase, the velocity field is $(\hbar/M)\,\nabla S$,
where $M$ is the mass of a particle of the fluid.
This relation requires that the circulation of the velocity 
field around any closed contour be quantized and equal to 
$\kappa \,h/M$, where $\kappa $ is an integer 
\cite{Onsager,Feynman}.  A non-zero value of $\kappa $ is 
associated with the presence of one or several quantum 
vortices.  Such a vortex is characterized by a line 
along which the condensate density vanishes and around which
the phase of the wave function varies by $2\kappa \pi$
(for recent reviews, see \cite{Dalfovo99,Fetter01}).

A reliable way to nucleate a quantum vortex in a dilute 
Bose-Einstein condensate is to place the system in a rotating
potential as was originally done with superfluid helium
in the famous ``rotating bucket experiment" 
\cite{Lifshitz,Donnelly}.  Starting with a condensate confined
in a cylindrically symmetric trap, one can stir the 
atomic cloud using an additional rotating anisotropic potential
created by the dipole potential of a non-resonant stirring
laser beam \cite{Madison00,Aboshaeer01}.
There exists a range of values of the stirring frequency for which 
one can nucleate in this way a single vortex which will center
itself on the symmetry axis of the trap. The presence of the 
vortex is then revealed by the existence of a density dip
at the center of the spatial distribution of the condensate. Further 
experiments have also revealed
that the angular momentum per particle is equal to $\hbar$
for this centered, one-vortex state \cite{Chevy00,Haljan01}. 

We present in this article a direct observation of the phase 
of such a rotating condensate.  We follow the scheme proposed 
in \cite{Bolda98,Tempere98,Castin99,Dobreck99}, which relies 
on an interferometric measurement.  Here, we divide the wave 
function of the condensate into two components separated in 
position and momentum. In the location where these components overlap,
a matter-wave interference fringe pattern is produced which
reveals the phase difference between the two components.
This homodyne detection of the phase pattern of the vortex
state of the condensate is analogous to the experiments 
aiming at studying the transverse 
phase structure of optical Laguerre-Gauss modes 
\cite{LaguerreGauss}.  Moreover, it is complementary 
to the heterodyne detection of a vortex state
performed in Boulder in 1999 \cite{Matthews99}.  In that
experiment, a vortex was created in a two component 
condensate by a direct printing of the vortex state 
phase $e^{i\theta}$ on one of the two components 
($\theta$ is the azimuthal angle around the symmetry axis 
of the condensate). The second component, corresponding to
a different internal state of the atoms, remained at rest 
at the center of the confining potential and was used as a
``reference'' matter wave to detect the phase of the vortex
state contained in the rotating component.  In this case,
the two interferring components correspond to two 
amplitudes with a different spatial phase
while in our case the two components have the same spatial 
phase but are shifted in position and momentum.

We prepare our Bose-Einstein condensate by evaporative 
cooling of $10^8$ rubidium ($^{87}$Rb) atoms confined in a 
Ioffe-Pritchard trap. 
The trap is axisymmetric and its eigenfrequencies 
are respectively $\omega_\perp=2\pi\times 192$~Hz and 
$\omega_z=2\pi\times 11.7$~Hz for the 
stretched state $m=+2$ of the $5\mathrm{S}_{1/2}, F=2$
ground state. The initial
temperature of the atom cloud, which is precooled using optical 
molasses,
is $100\;\mu$K. At the end of the evaporation, a condensate 
containing 
$3\;10^5$ atoms is formed, and the temperature of the cloud is below 
100~nK.
We then nucleate the vortices by the method described in 
\cite{Madison00}: we superimpose to the 
isotropic magnetic trap a far detuned laser beam propagating 
along the weak horizontal axis ($z$) of the trap and positioned 
by two crossed acousto-optic modulators. This light beam creates in 
the $xy$ plane
an anisotropic optical dipole potential of the form
$\delta U(X,Y)=\epsilon\, M\omega_\perp^2 (X^2-Y^2)/2$.
We rotate at constant frequency $\Omega/2\pi \sim 130$~Hz the eigenaxes $XY$ of 
this potential, which 
provides the desired stirring, with  $\epsilon=0.06$ in the present 
experiment. 
We stir the atomic cloud for a duration of 
300~ms. We verified in a preliminary step 
that this duration is sufficiently long to nucleate a well centered 
single vortex.

\begin{figure}[h]
\hskip -5mm
\epsfig{file=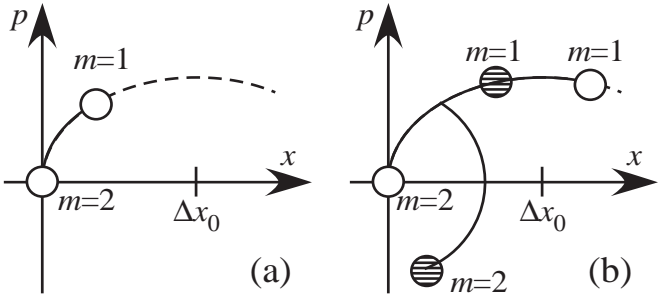,height=4.09cm,width=8.73cm}
\vskip 0.5cm
\caption{
Phase-space trajectories of the ${m=+2/m=+1}$ Zeeman substate
wave packets in the harmonic trapping potential during the 
experimental sequence. (a):  location of the two wave packets
at a time $\tau_1$ after the first rf pulse. 
(b): location of the four confined wave packets
at a time $\tau_2$ after the second rf pulse. 
The hatched disks correspond to the wave packets 
generated by the second rf pulse.}
\label{phasespace}
\end{figure}

To realize the amplitude splitting interferometer, we use a method 
similar to the one described in \cite{Inguscio01}.
A succession of two radio frequency (rf) pulses, 
which resonantly drive transitions between different internal states
of each rubidium atom, is used to coherently split and recombine 
the condensate cloud.
The atomic rubidium condensate is initially polarized the $F=m=2$ Zeeman substate.  
The first rf pulse couples this magnetic
spin state to the four other Zeeman substates of the $F=2$ manifold.
Among these four, only the $m=+1$ is also magnetically trapped. 
However, due to gravity, the center of its confining potential
is vertically displaced with respect to the center of the $m=+2$ trap
by an amount $\Delta x_0=g/\omega_\perp^2\approx 6~\mu$m.
After this first pulse, the atoms are allowed to evolve
in the harmonic trapping potential for a variable time $\tau_1$.
During this time, those atoms driven into the other Zeeman
substates fall or are ejected out of the trap. What 
remains are the two trapped clouds now with slightly 
different momenta and positions as determined by their
respective phase space trajectories pictured in
Fig.~\ref{phasespace}a.
When the second rf pulse is applied, each trapped 
state generates a copy in the other 
trapped Zeeman sublevel making a total of four wave packets 
present in the trap (two in $m=+1$ and two in $m=+2$). 
We then let the atoms evolve in the harmonic potential
for a time $\tau_2$ (Fig.~\ref{phasespace}b) after which we
extinguish the trapping potential and let the condensate clouds
freely expand and fall for 25~ms.  Finally, we record
the atomic interference pattern between the overlaping wave 
packets of a given internal Zeeman spin state 
using the resonant absorption imaging technique originally
used to detect the density dip associated with the vortex line.

The possibility of changing independently $\tau_1$ and 
$\tau_2$ allows for an independent control of the initial relative 
velocity and separation of two interfering wave packets. Note that 
the amplitude of these rf pulses is large enough so as to 
resonantly excite all atoms in the condensate, 
in spite of the inhomogeneity of the magnetic field inside the trap. 
This requires that the Rabi frequency $\Omega_{\rm rf}$ associated 
with these rf excitations
be at least of the order of $\mu/\hbar$, where $\mu$ is the chemical 
potential. 
In our case, $\mu/\hbar$ is typically  $2\pi\times 10$~kHz while 
$\Omega_{\rm rf}=2\pi\times 7$~kHz. 
The duration of each rf pulse is $13\;\mu$s which ensures that the 
populations of the two states $m=+2$ and $m=+1$ after each pulse 
are comparable.

\begin{figure}[ht]
\centerline{\epsfig{file=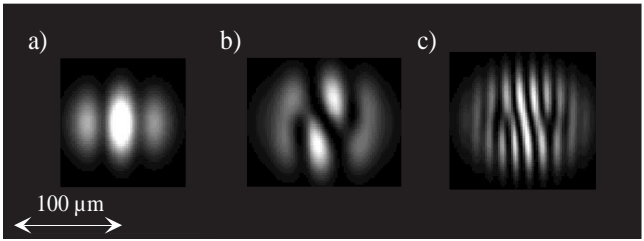,height=3.17cm,width=8.5cm}}
\vskip 0.5cm
\caption{Expected fringe pattern of a Bose-Einstein condensate 
initially splitted into two parts
and undergoing a free expansion phase. Figure (a) is without a vortex 
and (b,c) are with a vortex. 
For $(b)$, close to our experimental conditions, the fringe spacing 
$x_s$
is equal to 39 $\mu$m, 
and it is equal to the separation of the vortex cores after expansion 
$|{\bf r}_1-{\bf r'}_1|$. (c): same as (b), with a fringe spacing
$x_s=13\;\mu$m$=|{\bf r}_1-{\bf r'}_1|/3$ (this fringe spacing is too 
small to be detected 
in our experimental setup).
For figures (b) and (c), the relative phase of the two condensates is 
$\pi$.}
\label{fringetheo}
\end{figure}

Before showing the interference patterns detected experimentally,
we present a simple theoretical model to discuss the type of 
signal which can be expected. Consider first
a trapped condensate localized around ${\bf r}=0$ with no velocity 
field 
($\psi({\bf r})=\sqrt{n({\bf r})}$). We assume that this condensate
is split into two
parts separated by ${\bf a}$, so that the unnormalized state of the 
system is $\psi({\bf r}-{\bf a}/2) + \psi({\bf r}+{\bf a}/2)$. 
The potential confining the atoms is then 
extinguished and the two clouds overlap and interfere. The detection of 
the interference
pattern is done at a time $t$ large enough so that the final size of 
the 
condensate is much larger than the initial one. Consequently a model 
relying on a non-interacting 
Bose-Einstein condensate is sufficient to interprete the fringe 
pattern 
\cite{Castin96,Kagan97,meanfield}. The phase pattern of the wavefunction 
issued from the condensate initially
localized in ${\bf r_0}= \pm{\bf a}/2$ is the same as that of the function
\begin{equation}
\Phi({\bf r})=\exp \left(\frac{iM({\bf r}-{\bf r}_0)^2}{2\hbar t}  \right)
\label{phase0}
\end{equation}
so that the interference pattern consists in parallel fringes, with a 
fringe
spacing \cite{Andrews97}:
\begin{equation}
x_s=\frac{ht}{Ma}\ .
\label{interfrange}
\end{equation} 
The phase pattern of (\ref{phase0}) remains valid even 
if the cloud initially localized in ${\bf r}_0$ has a non zero 
mean velocity. Therefore the expression
for the fringe spacing (\ref{interfrange}) is still correct  
when the relative velocity of the two interfering clouds is not zero, 
provided $a$ denotes
the {\it initial} spatial separation between the centers of the 
condensates before
the atoms were released \cite{Inguscio01}. 
Note that the size of the expanding condensates also increases
linearly with $t$, so that the number of visible fringes is 
independent of the expansion  time.
In practice, the number of visible fringes is limited by the
optical resolution $\delta x$ of the imaging system ($\delta 
x=7\;\mu$m in our experiment),
since one must have $\delta x \ll x_s$. 
Choosing the maximal value of the expansion time 
compatible with our experimental setup (25~ms) gives
from (\ref{interfrange}) the product
$x_s a \sim (11\;\mu {\rm m})^2$. 
Then for a good fringe visibility, we choose
a small value of $a$, typically $a=2.5\;\mu$m, which yields $x_s \sim 
45\;\mu$m.
The diameter of the condensate after expansion is 100$\;\mu$m, so 
that 3 bright fringes
should be clearly visible in the interference pattern. The 
corresponding calculated 
pattern is showed in Fig.~\ref{fringetheo}a assuming an initial 
Gaussian distribution 
with an r.m.s. width $\Delta x=\Delta y=0.5\;\mu$m, a separation 
$a=2.9\;\mu$m and an expansion time of $25$~ms, yielding 
$x_s=39\;\mu$m.

Consider now a condensate centered at ${\bf r}={\bf r}_0$, and assume 
that a quantized
vortex, with an angular momentum $+\hbar$ with respect to the $z$ 
axis, 
sits at the center of the condensate. The initial wave function
of the condensate is $\psi({\bf r})=(x-x_0+i(y-y_0))\;f(|{\vec 
\rho}-{\vec \rho}_0|,z)\;e^{i{\bf k}\cdot{\bf r}}$.
Here we use cylindrical coordinates ($\vec \rho,z$) with respect to 
the $z$ axis, and $f$ 
is a function of $|\vec \rho|$ and $z$ with a constant phase.
When the condensate is released from the trap and expands 
for a time $t$, the phase pattern of the
cloud wavefunction is equal to that of the function $\Phi'$ 
defined as: 
\begin{equation}
\Phi'({\bf r})=
(x-x_1 + i(y-y_1))\;\exp\left( \frac{iM ({\bf r}-{\bf r}_0)^2}{2\hbar 
t} \right)
\label{phasepattern}
\end{equation}
where we set ${\bf r}_1={\bf r}_0+{\bf v}_0t$.  
Note that we allow for a possible non zero velocity 
${\bf v}_0=\hbar {\bf k}/M$ of the center of mass of the condensate. 
This result can be shown first for a cloud initially
at rest (${\bf v}_0=0$), and can then be generalized to a moving 
cloud using a Galilean transform.

The expected signal in an interference experiment can be obtained by 
superposing two
wavefunctions varying as (\ref{phasepattern}), obtained by expanding 
two condensates initially
located at ${\bf r}_0={\bf a}/2$ and ${\bf r}_0'=-{\bf a}/2$, with 
initial velocities ${\bf v}_0$
and ${\bf v}_0'$. The interference pattern consists in parallel lines 
separated by $x_s=ht/Ma$,
with ``defects" associated with the presence of the vortex. These 
defects  
are clearly visible if the distance ${\bf r}_1-{\bf r'}_1={\bf 
a}+({\bf v}_0-{\bf v'}_0)t$ 
between the two vortex cores after expansion is larger than $x_s$. 
Since $a$ is very small
compared to $x_s$ in our setup, this visibility condition requires
that:
\begin{equation}
\Delta v \geq \frac{h}{Ma}
\label{visibility}
\end{equation}
where $\Delta v=|{\bf v}_0-{\bf v'}_0|$.
On the other hand $\Delta v$ should be smaller than the velocity 
width of each condensate after expansion,
so that the two clouds overlap after the time-of-flight. In practice,
we choose the times $\tau_1$ and $\tau_2$ so that (\ref{visibility}) is close 
to being an equality and the overlap of the two wave packets
after expansion is maximized. 
The centers of the two vortices after expansion are then separated by 
the fringe 
spacing $x_s$ and one expects the characteristic fringe pattern shown 
in 
Fig.~\ref{fringetheo}b, with two dislocations (``H" shape) indicating 
the centers 
of the two vortices after time of flight.
In Fig.~\ref{fringetheo}c, we show the fringe pattern which would be 
expected for a larger
initial separation $a$, where more fringes are visible.

\begin{figure}[t]
\centerline{\epsfig{file=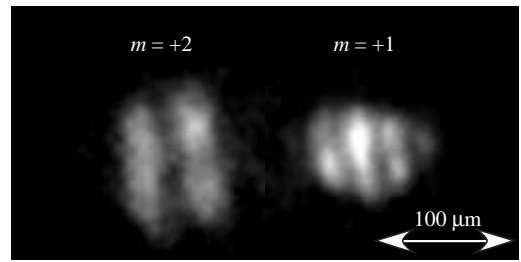,width=6.77cm,height=3.39cm}}
\vskip 0.5cm
\caption{Interference pattern of a non rotating 
condensate split into 4 parts (2 in each Zeeman state $m=+2$ and 
$m=1$), which overlap
during the free expansion phase. Here $\tau_1=0.3$~ms and 
$\tau_2=0.4$~ms.}
\label{fringeexp}
\end{figure}

We now turn to the discussion of the experimental results. We have 
first 
measured the interference pattern in the absence of any vortices. 
As expected,
this pattern consists in straight fringes, and a typical result is 
shown on 
Fig.~\ref{fringeexp}. The two patterns corresponding to the $m=+2$ 
and $m=+1$ components are recorded simultaneously. 
The spatial separation between these two components results from 
the difference between the initial velocities of the $m=+1$ and
$m=+2$ clouds when released from the trap. The difference in the 
fringe spacings reflects the fact that the initial separation $a_1$
between the two wavepackets in state $m=+1$ was larger than  
the separation $a_2$ between the two wavepackets in $m=+2$ (see e.g. Fig.
\ref{phasespace}).
We have checked for each component 
that the variation of the fringe spacing with $a_{1,2}$ 
is in good agreement with 
(\ref{interfrange})
(see also \cite{Inguscio01}). In the following we use
the interference pattern obtained with the $m=2$ component, which offers
a better visibility due to its larger spatial extent after time of flight,
which is itself due to a stronger confinement in the trap. 

\begin{figure}[t]
\centerline{\epsfig{file=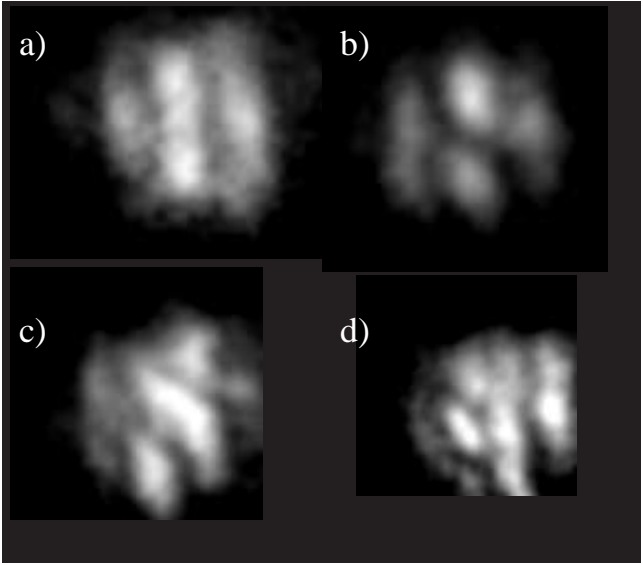,height=7.51cm,width=8.5cm}}
\vskip 0.5cm
\caption{Interference pattern measured in the $m$=+2 channel 
with no (a), one (b-c) and several (d) vortices. For these 
pictures, $\tau_1=0.688~$ms and $\tau_2=1.320$~ms. The
stirring frequency was set to  $\Omega=2\pi\times 125$~Hz (a),
$\Omega=2\pi\times 130$~Hz (b-c) and $\Omega=2\pi\times 154$~Hz
(d). The patterns (b) and (c) were recorded with the same initial
conditions, and the change in the interference pattern
results from a change in the relative phase of the two parts of the 
condensate.}
\label{franges}
\end{figure}

When we stir the atomic cloud at a frequency
of 130~Hz (close to the critical frequency 
$\omega_\perp/\sqrt{2}$ \cite{Madison01}), 
a well centered, single vortex is created 
in the condensate. The result from an 
interference experiment performed
in this case is shown on Fig.~\ref{franges}b for the $m=+2$ 
channel.  This pattern, which is very different from the one 
obtained with a slightly slower rotation (125~Hz)
at which no vortex nucleation occurs 
(Fig.~\ref{franges}a), is the main result of this paper. 
It clearly shows the telltale ``H" shape characteristic 
of the $2 \pi$ phase winding due to the
presence of a vortex (see Fig.~\ref{fringetheo}b). 
We repeated this experiment many times for the same initial 
condition, and have found that the interference patterns
produced have a similar shape but with a relative phase 
between the two components which fluctuates from
one shot to another so that the bright and dark regions in 
Fig.~\ref{franges}b
can be exchanged (see e.g. Fig.~\ref{franges}c). 
Finally, when several vortices are nucleated by stirring at a higher 
frequency, the fringe structure is more complex 
and the number of fringe dislocations is larger 
(Fig.~\ref{franges}d).

In conclusion, we have observed the $2\pi$ phase shift 
associated with the presence of a vortex 
in a single component Bose-Einstein condensate 
by using an interferometric detection method. 
This technique can be used as a very sensitive tool
to diagnose the presence of vortices generated by the fast motion
of a macroscopic object in a condensate. This was demonstrated very 
recently at MIT, in an experiment where the interference
pattern between a condensate with vortices and a condensate at rest
has been recorded \cite{Inouye01}. 

The ``beam splitter" used here is very simple, since it consists
of a resonant radio frequency pulse. 
However, one disadvantage of this type of ``beam splitter''
is the unavoidable loss of atoms to the other untrapped Zeeman 
states resulting in at most only $\sim1/3$ of the atoms 
participating in the interference signal in the $m=+2$ channel. 
We plan to use in the near future a more efficient scheme
discussed in \cite{Dobreck99}, which involves the use of two 
Bragg pulses to couple just two states of different momentum.

{\acknowledgments
We thank the ENS Laser cooling
group for helpful discussions. 
This work was partially supported by CNRS, Coll\`{e}ge de France,
DRET, and DRED.  K.M. acknowledges DEPHY for support.


\begin{references}
\bibitem{meanfield} More precisely, if the two parts of the 
condensate are initally well separated, 
the interaction between these parts is negligible, since the density 
is already very low when they
start to overlap. The only effect of the mean field interaction 
(within each 
part) is a modification of the envelope of the interference pattern.
\end{references}
\end{document}